\documentclass[pre,10pt]{revtex4-2}

\usepackage{graphicx}
\usepackage{amsmath}
\usepackage[utf8]{inputenc}
\usepackage{hyperref}
\usepackage{url}
\usepackage{amssymb}
\usepackage{amsfonts}
\usepackage{indentfirst}
\usepackage[toc,title]{appendix}
\usepackage{color}
\usepackage{bm}
\usepackage{natbib}
\newcommand{\tr}{\textup{Tr}}

\newcommand{\<}{\left<}
\renewcommand{\>}{\right>}

\newcommand{\bb}[1]{{\bf #1}}
\newcommand{\re}{\textup{Re}}

\begin{document}

\title{Transient amplification in balanced neural networks}
\author{ Wojciech Tarnowski}
\email{wojciech.tarnowski@doctoral.uj.edu.pl}
\affiliation{Institute of Theoretical Physics, Jagiellonian University, 30-348 Cracow, Poland}
\date{\today}

\begin{abstract}
Transient amplification has been proposed as an important mechanism not only in neuroscience but in many areas modeled by dynamical systems. Despite that, there is no clear biologically plausible mechanism which fine-tunes the coupling matrix or selects signals to be amplified. In this work we quantitatively study transient dynamics in the Rajan-Abbott model of a recurrent neural network [K. Rajan and L.F. Abbot PRL 97, 188104 (2006)]. We find a second order transition between a phase of weakly or no amplified transients and a phase of strong amplification, where the average trajectory is amplified. In the latter phase the combination of Dale's principle and excitatory/inhibitory balance allows for strong weights, while maintaining the system at the edge of chaos. Moreover, we show that the amplification goes hand in hand with greater variability of the dynamics. By numerically studying the full probability density of the squared norm, we observe as the strength of weights grows, the right tail of the distribution becomes heavier, moving from the Gaussian to the exponential tail.

\end{abstract}

%\abstract{aaa}
\maketitle

\section{Introduction}

Dynamical systems are common tools in modeling complex systems. They exhibit a variety of possible long-time behavior including fixed points, limit cycles, synchronization and chaos~\cite{Strogatz2018}. Before a system reaches an asymptotic state, its early-time dynamics can be completely different than the long-time behavior. This is called the transient phase. 
Due to the nonlinear interactions between components the study of the temporal behavior is often limited to numerical simulations. More can be said in the vicinity of a stable fixed point, where dynamics can be linearized. Despite the significant reduction of complexity, the linear model with a non-normal coupling matrix can exhibit non-trivial behavior. When such a system is slightly perturbed, some trajectories may initially drift away from the fixed point, sometimes invalidating the linear approximation~\cite{Trefethen,Trefethen2,ChaosTransition}. This counter-intuitive behavior is caused by non-orthogonal eigenvectors~\cite{Trefethen2,GrelaTransient}, which amplify perturbations.

Non-normality is ubiquitous in complex systems~\cite{Carletti}, since its simplest source is the asymmetry of interactions between components. One therefore expects transient dynamics to be a prevailing phenomenon. Indeed, the transient amplification of small noise-induced perturbations was recently proposed as a mechanism of formation of Turing patterns~\cite{Turing1,Turing2,Turing3}.
It has also been suggested that novel phenomena like the state consisting of clusters of phase-locked Kuramoto oscillators, dubbed \textit{chimera state}~\cite{Chimera0,Chimera01} or the many-body localization~\cite{MBL1,MBL2} are long living transients~\cite{Chimera,MBL}. Transients are key to understanding how ecological systems recover after sudden changes of environmental conditions~\cite{Eco1,Eco2,Eco3,Eco4,Eco5,Eco6,Eco7}. They play an important role also in gene regulatory networks~\cite{Gene1,Gene2,Gene3,Gene4}, epidemic spreading~\cite{Epid1,Epid2,Epid3} and the stability of complex networks~\cite{Netw1}. The analysis of propagating transients upon controlled perturbation is a useful tool in determining the network structure~\cite{Netw2,Netw3, Netw4, Netw5}. In neuroscience transient dynamics was proposed as a mechanism of amplification of weak neural signals~\cite{Neuro1,Neuro2,Neuro3} and a way of coding information~\cite{CodingTransient}. Due to transient dynamics networks with non-normal connectivity structure are more robust to noise and transmit information more efficiently~\cite{HennequinScience}.
In artificial recurrent neural networks non-normality of the connectivity matrix enhances expressivity and allows the network to learn long-time dependencies~\cite{ML}.

Besides the coupling matrix, the dynamics strongly depends on the specific perturbation applied, or, equivalently, on the initial condition. One therefore needs to consider ensembles of initial conditions and coupling matrices. In the absence of any prior information other than the norm of the perturbation, according to the Jaynes principle~\cite{Jaynes1,Jaynes2}, initial conditions are generated with the uniform measure on the hypersphere, which maximizes the information entropy. The choice of the coupling matrix can be addressed within the Random Matrix Theory, which provides the notion of a generic matrix as random.
A convenient object to quantitatively describe the transient dynamics is the squared Euclidean distance from the fixed point.  Within the above framework it was shown that for asymmetric random matrices the squared Euclidean distance takes a universal form, which is monotonic~\cite{ChalkerMehlig,MartiBrunelOstojic,Erdos2019,NeriVivo}. This means that the average trajectory is not transient. Moreover, only $9\%$ of eigendirections leads to transient amplification~\cite{CodingTransient}. 

If transients are of any importance in neuroscience, there must be a mechanism for either selecting lower dimensional space of perturbations or for fine-tuning the coupling matrix.  Hennequinn, Vogels and Gerstner~\cite{Neuro2} proposed an algorithm for fine-tuning by minimizing the spectral abscissa via gradient descent. Despite the fact that this model exhibits strong transient behavior and its dynamics is consistent with experimental data, the fine-tuning mechanism is not biologically sound. Rajan and Abbott proposed a more biologically motivated model~\cite{RajanAbbott} in which the strong eigenvector non-orthogonality was recently   observed numerically~\cite{Gudowska2020}.  In this work, we perform an analytical study of the transient dynamics in the Rajan-Abbott model.

\subsection{Setting}

We consider a dynamical system evolving according to the system of $N$ coupled first-order differential equations $\dot{x}_i=f_i(\bb{x})+\xi_i(t)$. Here $\bm{\xi}(t)$ represents an external driving of the system. In the context of neural networks  it is an external stimulation. Close to a fixed point $\bm{x^*}$ satisfying $\bm{f}(\bm{x}^*)=0$ the dynamics can be linearized to
\begin{equation}
\frac{d y_i(t)}{dt}=\sum_{k=1}^{N}J_{ik}y_k(t) +\xi_i(t), \label{eq:DiffEq}
\end{equation}
where $\bm{y}(t)=\bm{x}(t)-\bm{x^*}$ and $J_{ik}=\left.\frac{\partial f_i}{ \partial x_k}\right|_{\bm{x^*}}$ is the Jacobian.
%Long-time dynamics of the system is governed by the rightmost eigenvalue $\lambda_R$ of $J$ (i.e. the one with the largest real part) and according to the Hartman-Grobman theorem~\cite{Grobman,Hartman} the system is stable if $\textup{Re}(\lambda_R)<0$, i.e. all trajectories are attracted to a fixed point.  This characterization, however, does not say anything about the early-time evolution of \eqref{eq:DiffEq}. {\color{red} A nice figure summarizing this behavior?}
Here we consider a delta-pulse external driving $\bm{\xi}(t)=\delta(t)\bm{y_0}$, which corresponds to a single spike. The formal solution of \eqref{eq:DiffEq} reads
\begin{equation}
\bm{y}(t)=\exp\left( Jt\right) \bm{y_0}. \label{eq:FormalSol}
\end{equation}
The squared Euclidean distance from the fixed point is simply the squared norm of the solution 
\begin{equation}
||\bm{y}(t)||^2=\bm{y_0}^T e^{J^{T}t} e^{Jt} \bm{y_0}. \label{eq:Norm}
\end{equation}

\subsection{The role of non-orthogonal eigenvectors \label{sec:RoleEigenvectors}}

The eigenbasis of the coupling matrix, at which the time dynamic decouples, is different from the coordinate system at which the system is observed. Transient dynamics stems from the fact that the transformation between them is not an isometry, which is the case only if $J$ is normal.

If a non-normal matrix is diagonalizable, it possesses distinct left and right eigenvectors, satisfying the eigenproblems $J\bm{r}_j=\lambda_j\bm{r}_j$ and $J^T\bm{l}_j=\lambda_j \bm{l}_j$. They are not orthogonal among themselves $\bm{l}_i^{\dagger}\bm{l}_j\neq\delta_{ij}\neq\bm{r}_i^{\dagger}\bm{r}_j$ but normalized to $\bm{l}_i^{\dagger}\bm{r}_j=\delta_{ij}$. For normal matrices these sets of eigenvectors coalesce, $\bm{l}_j=\bm{r}_j$.

The spectral decomposition $J=\sum_{k=1}^{N}\lambda_k \bm{r}_k \bm{l}_k^{\dagger}$ offers a more precise view on the interplay between eigenbases, allowing one to rewrite the squared norm as 
\begin{equation}
||\bm{y}(t)||^2=\sum_{j,k=1}^{N}(\bm{y_0}^T\bm{l}_j)(\bm{r}_j^{\dagger}\bm{r}_k)(\bm{l}_k^{\dagger}\bm{y_0}) e^{t(\lambda_k+\bar{\lambda}_j)}, \label{eq:NormEigendecomposition}
\end{equation}
which for normal matrices ($\bm{r}_j^{\dagger}\bm{r}_k=\delta_{jk}$) simplifies to
\begin{equation}
||\bm{y}(t)||^2=\sum_{k=1}^{N}|\bm{y_0}^T\bm{l}_k|^2 e^{2t \re\lambda_k}. \label{eq:NormNormal1}
\end{equation}

The dynamics driven by normal matrices reduces to eigenspaces in which eigenmodes are decoupled. In stable systems the decay of the squared norm is driven only by real parts of eigenvalues. 
Due to non-orthogonality of eigenvectors in the non-normal case, all eigenmodes are coupled to each other, and the imaginary part of the eigenvalues produces an oscillatory behavior of the norm. The scalar products $\bm{r}_j^{\dagger}\bm{r}_k$ are not bounded, thus they may  amplify oscillations. The ascending oscillation amplified by the eigenvector non-orthogonality may overcome the exponential damping, resulting in the transient growth of the norm.

\subsection{Rajan-Abbott model}
The linear model \eqref{eq:DiffEq} is obtained by the linearization of the seminal model introduced by Crisanti, Sommers and Sompolinsky~\cite{NeuralNetworks2}. In its original form, the Jacobian is given as $J=-\mu + X$, where $X$ is a real Gaussian random matrix from the Ginibre ensemble. In the absence of the couplings, $X = 0$, the dynamics quickly ends at the fixed point. It was observed that  in the full nonlinear model, as the coupling strength increases, the system moves to a state with chaotic dynamics. This takes place precisely when the spectral radius of $X$ exceeds $\mu$, making the fixed point unstable.

Rajan and Abbott~\cite{RajanAbbott} introduced additional deterministic structure in the coupling matrix, modeling basic biological observations. 
In their model there are two types of neurons - excitatory and inhibitory. The strength of the interaction of the $j$-th neuron is a Gaussian random number of mean $\frac{m_j}{\sqrt{N}}$ and variance $\frac{1}{N}$, with $m_j>0$ for excitatory neurons and $m_j<0$ for inhibitory neurons, reflecting Dale's principle.
It is convenient to decompose the synaptic strength matrix as $X=M+Y$ where $M$ is a deterministic part, while random $Y$ captures fluctuations around the average. The elements of $Y$ are independent identically distributed random Gaussian numbers of zero mean and $1/N$ variance. It is a real Ginibre matrix~\cite{Ginibre}.
It was observed experimentally that the sum of excitations and inhibitions exerted at each neuron is maintained 0 on the time scale of milliseconds, the so-called excitatory/inhibitory (E/I) balance~\cite{EIbalance1,EIbalance2}. This is encapsulated in the model both on the level of averages, $\sum_{j=1}^{N}m_{j}=0$, and also for each specific neuron separately, $\sum_{j=1}^{N}Y_{ij}=0$. The E/I balance implies that the spectrum of $X$ is the same as that of $Y$, making it insensitive to the deterministic connections~\cite{RajanAbbott}.

\section{Results}

We fix the norm of the initial perturbation to $1$.
For a fixed matrix $J$ the mean and variance of the squared norm of the solution \eqref{eq:Norm} taken with respect to all initial conditions with unit norm, are given by 
\begin{align}
\overline{||\bm{y}(t)||^2} & =   \frac{1}{N}\tr\, e^{J^T t}e^{Jt}, \label{eq:mean} \\
\mbox{var}\left(||\bm{y}(t)||^2 \right) &= \frac{2}{N+2}\left(\frac{1}{N}\tr\, e^{J^Tt}e^{Jt}e^{J^Tt}e^{Jt}-\left(\overline{||\bm{y}(t)||^2}\right)^2\right). \label{eq:variance}
\end{align}
For typical matrices, the normalized trace does not grow with the size of the system~\cite{sinai1998,Augeri}, thus the variance vanishes at large $N$. 
Most of the trajectories evolve like the mean trajectory, so the mean squared norm is a good scalar quantity for the study of possible transient dynamics.

In many models, like the Rajan-Abbott model, the matrix $J$ has some randomness incorporated into it. We define $S(t) = \<\overline{||\bm{y}(t)||^2}\>$ and $\Sigma(t) = \<\mbox{var}\left(||\bm{y}(t)||^2 \right) \>$, the mean and variance of the squared norm averaged also over the randomness in $J$ and further refer to them simply as mean and variance. While formulas (\ref{eq:mean}-\ref{eq:variance}) are valid for any matrix size, we calculate expectations of normalized traces only in the limit $N\to\infty$.

In the Rajan-Abbott model the mean squared norm reads
\begin{equation}
S(t) = e^{-2\mu t}\left(I_0(2t) +F(I_0(2t)-1)\right), \label{eq:RAmean}
\end{equation}
while the variance in the leading order is
\begin{equation}
\Sigma(t) = 2F^2 e^{-4\mu t} (I_0(2t)-1)^2.     \label{eq:RAvar}
\end{equation}
 Here a single parameter $F = \frac{1}{N}\sum_{j=1}^{N}m_j^2 = \frac{1}{N} \tr MM^T$ captures the entire dependence on the deterministic connections.  Note also that despite \eqref{eq:variance} predicting $\Sigma(t)$ to scale as $1/N$, the variance in the Rajan-Abbott model does not vanish in the large $N$ limit, because the coupling matrix is atypical. This peculiarity originates from deterministic weights obeying Dale's principle and the E/I balance condition.

The time dynamics of the system is determined by both the eigenvalues and the eigenvectors of the coupling matrix. With the E/I balance  imposed the deterministic matrix $M$ does not influence the spectrum, offering a chance to separate the effect of non-orthogonal eigenvectors from the distribution of the eigenvalues on the transient dynamics. The results (\ref{eq:RAmean}-\ref{eq:RAvar}) should be contrasted with the analogous dynamics which would take place in the absence of the deterministic part, $F=0$. In such a case \eqref{eq:RAmean} reduces to $S(t) = \exp(-2\mu t) I_0(2t)$, in agreement with previous results~\cite{ChalkerMehlig,Erdos2019,NeriVivo}. The variance is of smaller order in $N$ and reads
\begin{equation}
\Sigma(t) = \frac{2e^{-4\mu t}}{N}\left[ I_0(4t)+tI_1(4t)-2t I_1(2t)I_0(2t)-I_0^2(2t)\right].
\end{equation} 
In the absence of the deterministic matrix trajectories concentrate around the mean trajectory.

There is much evidence that neural systems operate in a critical regime between chaos and order~\cite{Dante,Criticality2}, so in the further analysis we tune the system to criticality by setting $\mu=1$.

\begin{figure}\begin{center}
\includegraphics[width=0.33\textwidth]{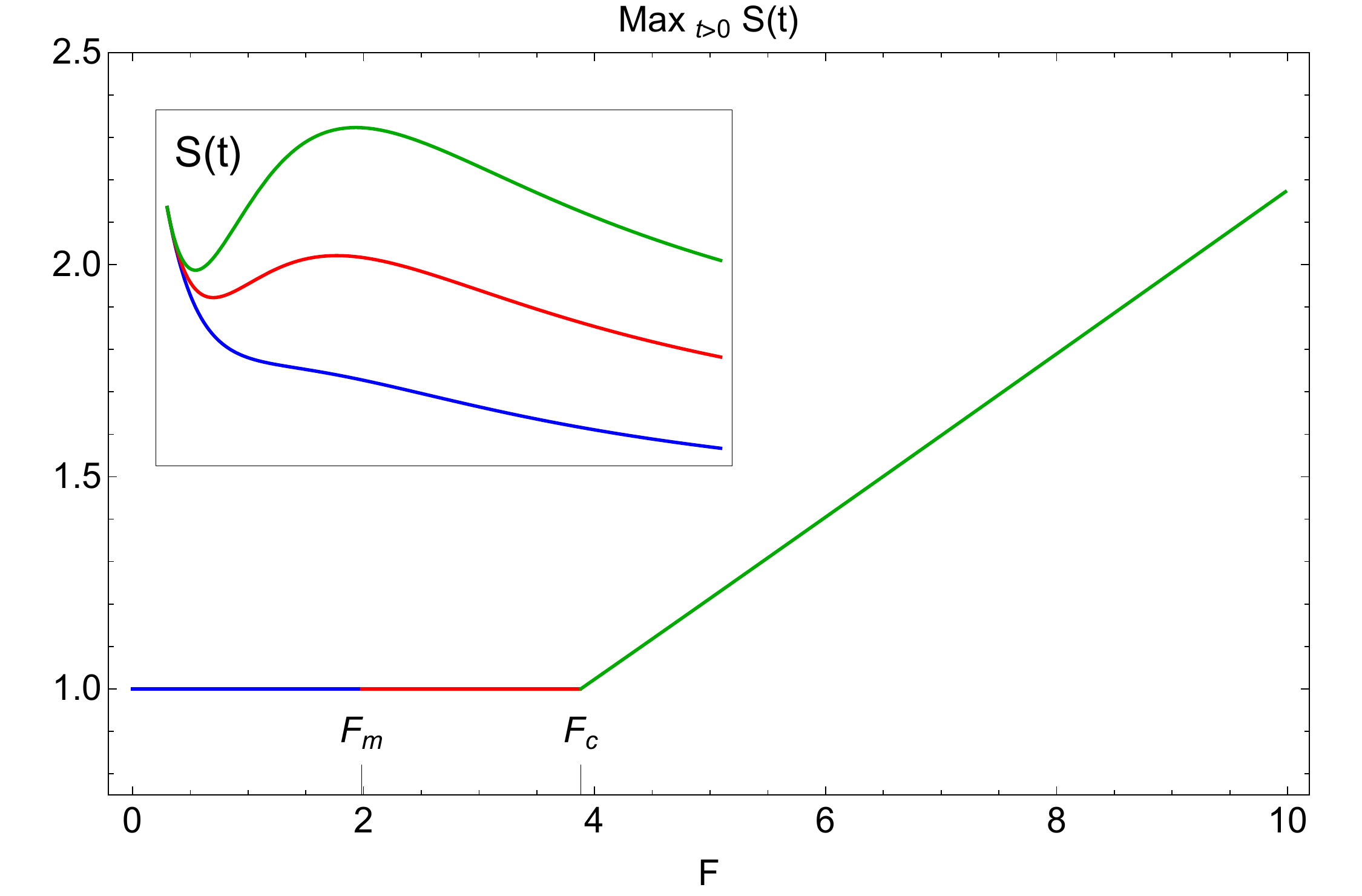}\includegraphics[width=0.33\textwidth]{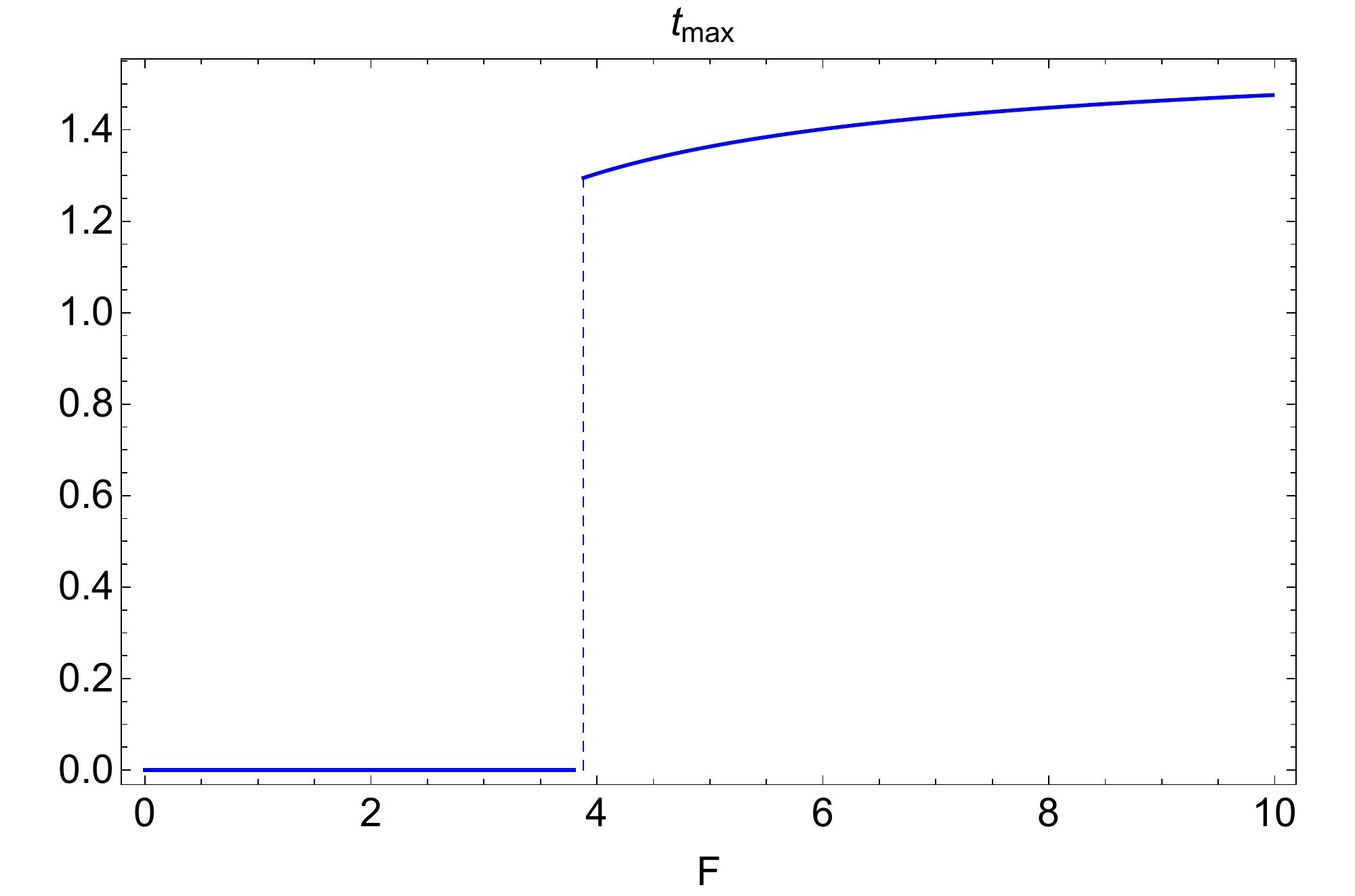}\includegraphics[width=0.33\textwidth]{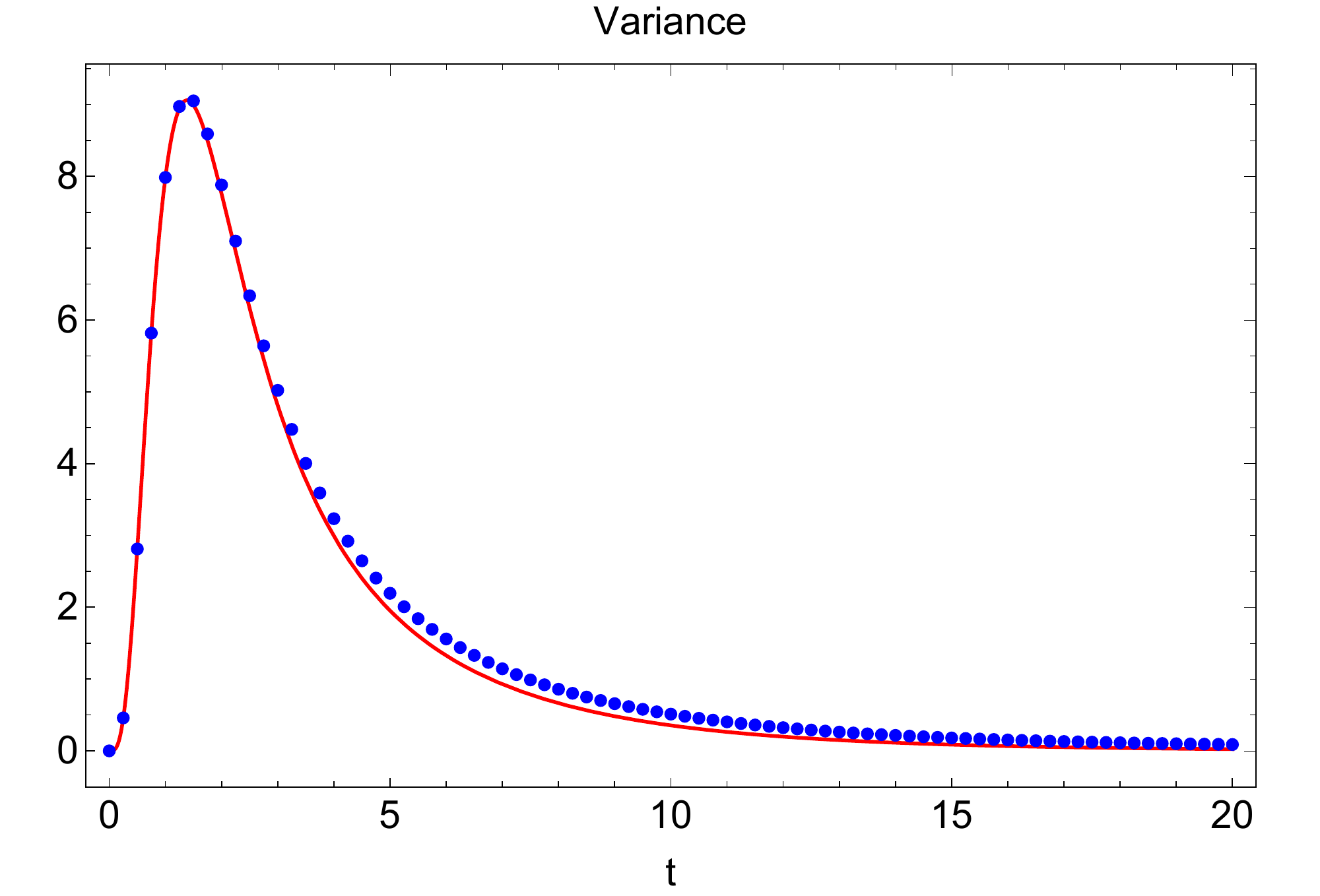}
\end{center}
\caption{(left) A phase diagram of the transient dynamics, depicting the maximum value of the average squared norm. For $0\leq F \leq F_m$ $S(t)$ decays monotonically. For $F>F_m$, transient amplification causes $S(t)$ to develop a local maximum, but only for $F>F_c$ the value at this maximum is greater than the initial value. The inset presents a an average trajectory in each regime. (middle) Dependence of the time at which $S(t)$ attains its maximum on the parameter $F$. There is a discontinuous transition at $F_c$. (right) Variance \eqref{eq:RAvar} compared with the numerical simulations of the dynamics in the Rajan-Abbott model. We used the weight matrix of size $N=500$ with a fraction of $0.85$ excitatory neurons with $m_E=1.5$ and $0.15$ inhibitory neurons with $m_I=-8.5$. For a better numerical stability we set $\mu=1.05$. Each point corresponds to averaging over 500 realization of randomness in the coupling matrix of size $N=500$ and 500 initial conditions for each matrix. 
\label{Fig1}
}
\end{figure}

To quantitatively address the effect of deterministic connection on the transient amplification, it is convenient to consider the maximum of $S(t)$ attained over the entire evolution, $\max_{t>0}S(t)$.
Analyzing the dependence of $S(t)$ on the parameter $F$, we find that for $F\leq F_{m}=1.9835$ the average squared norm decays monotonically. At $F_m$ it starts developing a local maximum, demonstrating the significant effect of transient amplification. At this regime, transient amplification is weak and $S(t)$ at this local maximum does not exceed the initial value. Only after $F_c=3.8813$ the value at the maximum is greater than 1 and grows linearly with $F$, see Fig.~\ref{Fig1}. At the critical value $F_c$ the system therefore undergoes a second order transition to a phase dominated by the transiently amplified dynamics. Analogously, defining $t_{max} = \mbox{arg max}_{t\geq 0} S(t)$, one observes a transition at $F_c$, but this time discontinuous. Beyond $F_c$ one observes further slow growth of $t_{max}$. The effect of the deterministic matrix on the transient dynamics is therefore twofold: it not only increases the magnitude of the amplification, but also extends the duration of this phase. 

It needs to be emphasized that the calculation of $\Sigma(t)$ is valid for large $N$ and relies on additional assumptions, thus the validity of~\eqref{eq:RAvar} needs to be verified a posteriori. The results of numerical simulation presented in Fig.~\ref{Fig1} show a good agreement. Variance of the squared norm has a single maximum at $t=1.586$, which does not depend on $F$. Its value is only slightly larger than $t_{max}$. 

The above results are valid up to a certain time $t^*(N)$, after which numerical results deviate exponentially with time. This behavior is caused by the fluctuations of the rightmost eigenvalue and corresponds to switching from the transient to the asymptotic regime~\cite{NeriVivo}.

\begin{figure}
\includegraphics[width=0.49\textwidth]{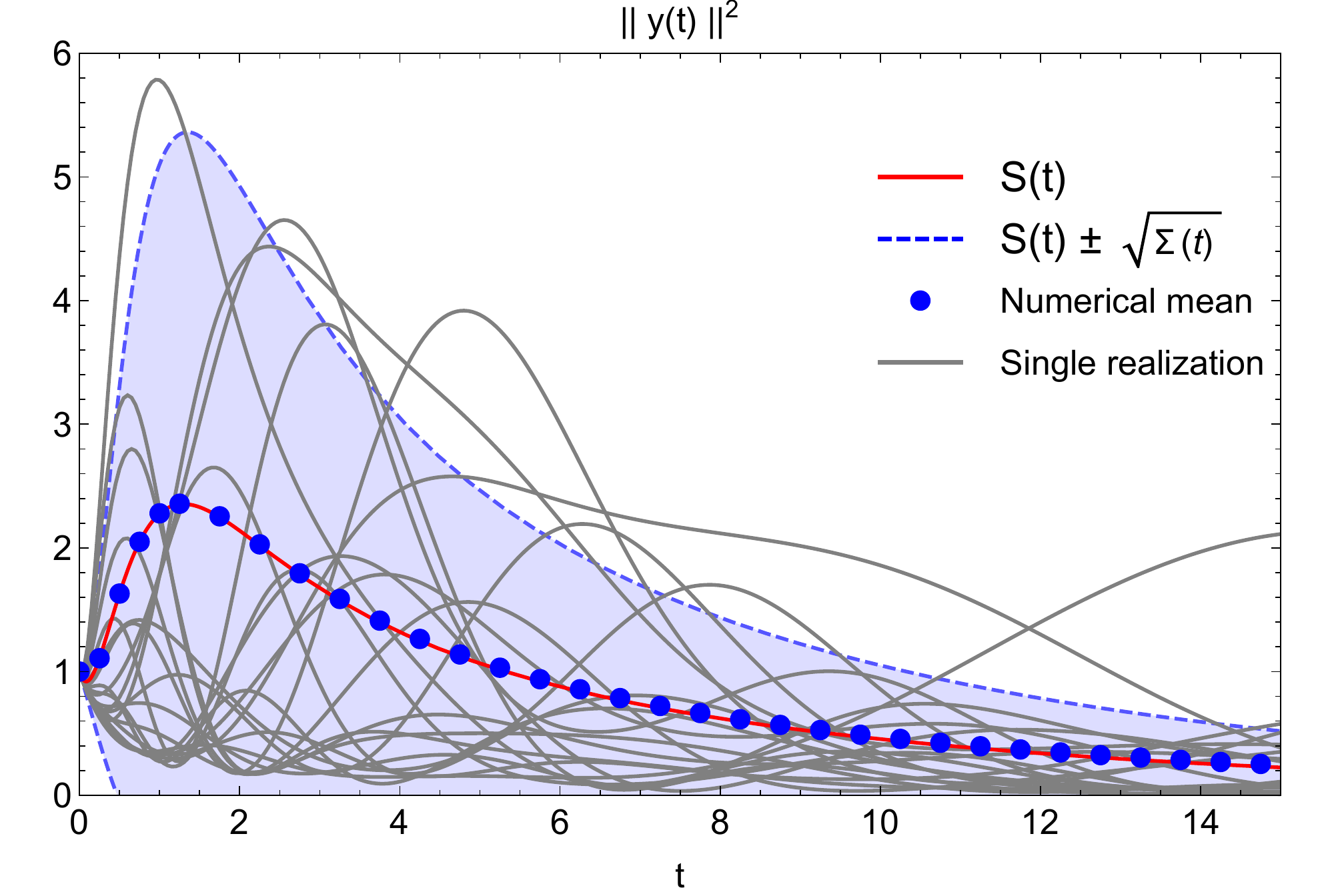} \includegraphics[width=0.5\textwidth]{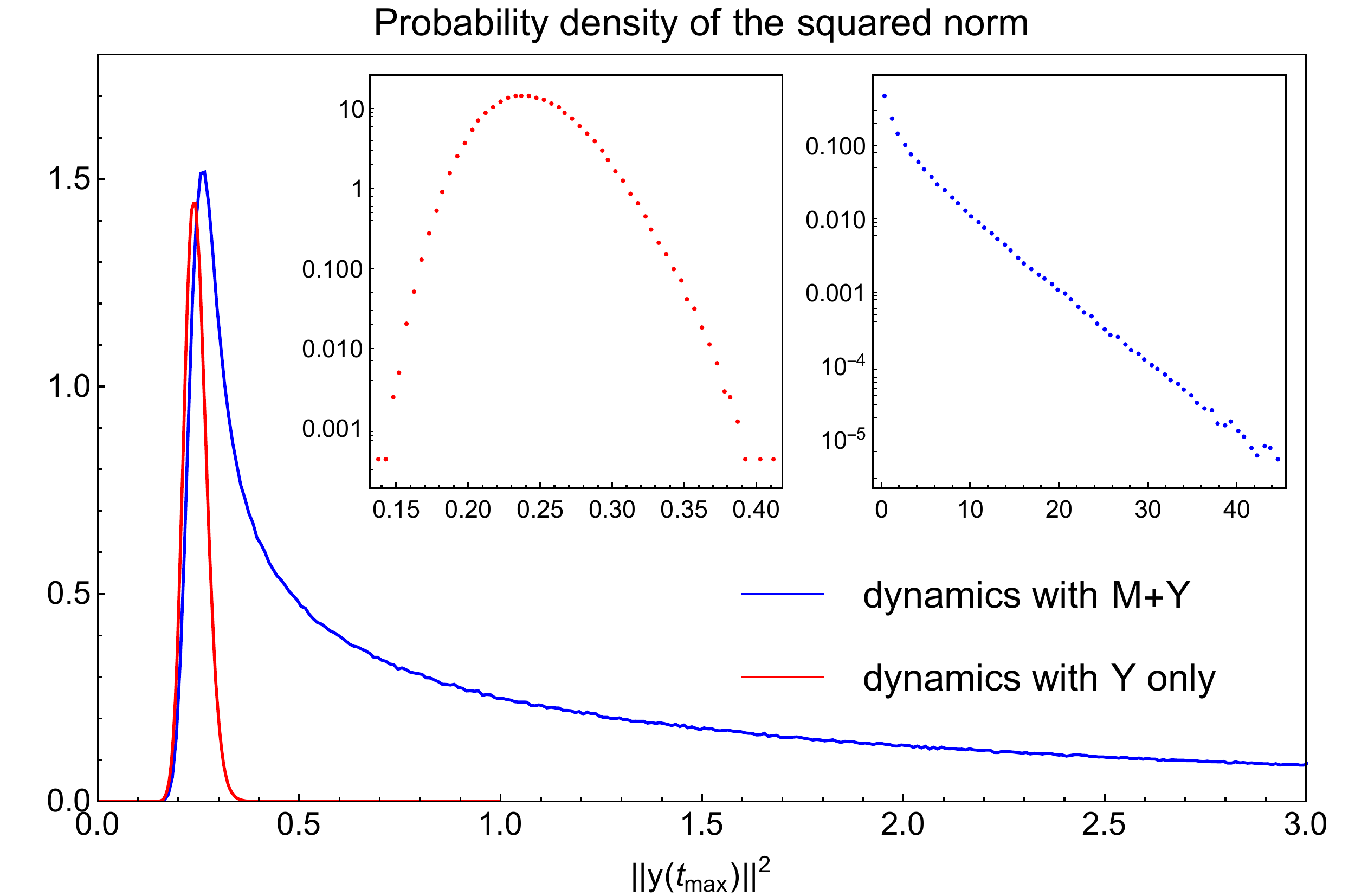}
\caption{(left) Numerical simulation of the squared norm in the Rajan-Abbott model. We used the same parameters as in Fig.~\ref{Fig1}. Solid gray lines represent 20 individual trajectories starting from random initial condition with unit norm, blue dots are numerical averages over randomness and initial conditions. Red solid line represents formula~\eqref{eq:RAmean}, while blue dashed line bound the shaded region corresponding to 1 standard deviation from the mean, calculated from~\eqref{eq:RAvar}.
(right) Full probability density of the squared norm at $t=t_{max}$. We juxtapose squared norms in (blue) the Rajan-Abbott model and (red) the model corresponding to the absence of deterministic weights. The latter density is downscaled by a factor of 10 to fit the image. In the insets we show these distributions on the log scale to emphasize the difference between their tails. We used the same deterministic weight matrix as in Fig.~\ref{Fig1}, but we set $\mu=1$.
\label{Fig2}
}
\end{figure}

The presented analytic approach is based on first two cumulants of the full distribution of the squared norm, which at this moment is not accessible analytically. We therefore resort to the numerical study of the probability density of $||\bm{y}(t)||^2$. In Fig.~\ref{Fig2} we present the distribution of the squared norm at $t=t_{max}$ and juxtapose it with the same quantity but calculated in the absence of the deterministic connections. The latter tightly concentrates around its mean value, as predicted by~\eqref{eq:variance}. Interestingly, modal values of both distributions are close to each other, but the distributions strongly differ in their right tails. In the presence of the deterministic connections the distribution develops an exponential tail, which shifts mean and significantly increases variance.

\section{Discussion} 
 
The squared norm is a good quantity for the analysis of the transient dynamics. 
In typical systems the full distribution of the squared norm tightly concentrates around its mean, with variance vanishing with the systems size. The average squared norm decays monotonically with time, only a tiny part of initial conditions is amplified. This therefore raises a question about a mechanism for either selection of proper initial conditions or fine-tuning the coupling matrix, making it atypical.
 
 In theoretical neuroscience a possible mechanism of fine-tuning comes from an interplay of Dale's principle and E/I balance. It not only it ties eigenvalues to the disk, preventing this way a transition to chaos~\cite{RajanAbbott}, but also causes strong eigenvector non-orthogonality~\cite{Gudowska2020}. A model incorporating these two effects allows also for an analytic treatment.
 
The dynamics in the model depends on a single parameter proportional to the square of weight strength. When it reaches critical value, trajectories are amplified on average, thus there is no need for selecting initial conditions. For a more quantitative treatment, one considers the maximum of the average squared norm through the entire evolution as a measure of amplification. In this variable one observes a second order transition between a phase of weak or no amplification to the strongly amplified transients. The variance of the squared norm initially grows with time, reaching its maximum slightly later than the maximum of the mean is reached. This shows that with the maximum average amplification comes also greater variability of trajectories, suggesting that in this model robustness against noise comes hand in hand with high information capacity. Numerical study of the full distribution demonstrates that the increase of the strength of connections makes the right tail of the distribution heavier from the Gaussian to the exponential one, leaving the modal value almost intact.

\section{Methods}
\subsection{Randomness in initial conditions}

The squared norm is a quadratic form in components of the initial vector $\bm{y_0}$. Having this dependence in mind, we further assume that the initial condition has a unit norm. The manifold of all initial conditions with fixed norm is a $N$-dimensional sphere, which is a compact set. Therefore, we probe initial conditions with a uniform density on the $N$-sphere, which maximizes the information entropy. 
This is realized by choosing a single representative vector of unit norm and generating all other initial conditions by the action of matrices generated from the Haar measure on $O(N)$, which is a group of isometries of $N$-sphere. The mean and  variance of $||\bm{y}(t)||^2$ with respect to initial conditions can be calculated by averaging over $O(N)$. 

Let $\bm{\psi}$ denote this representative vector and $A=e^{J^Tt}e^{Jt}$. Then $||\bm{y}(t)||^2 =\sum_{i,j,k,l=1}^{N}\psi_i O_{ij}A_{jk}O_{lk}\psi_l$. The expectations over orthogonal matrices are given by~\cite[Lemma 9]{ChatterjeeMeckes}
 \begin{equation}
 \overline{O_{ij}O_{\alpha\beta}}=\frac{1}{N}\delta_{i\alpha}\delta_{j\beta}, \label{eq:O1av}
 \end{equation}
 \begin{eqnarray}
 \overline{O_{ij}O_{rs}O_{\alpha\beta}O_{\lambda\mu}}=\frac{-1}{(N-1)N(N+2)}\left[\delta_{ir}\delta_{\alpha\lambda}\delta_{j\beta}\delta_{s\mu}+\delta_{ir}\delta_{\alpha\lambda}\delta_{j\mu}\delta_{s\beta}+\delta_{i\alpha}\delta_{r\lambda}\delta_{js}\delta_{\beta\mu}+ \delta_{i\alpha}\delta_{r\lambda}\delta_{j\mu}\delta_{\beta s}\right. \nonumber \\
 +\delta_{i\lambda}\delta_{r\alpha}\delta_{js}\delta_{\beta\mu}
 \left. +\delta_{i\lambda}\delta_{r\alpha}\delta_{j\beta}\delta_{s\mu}\right]+ \frac{N+1}{(N-1)N(N+2)}\left[\delta_{ir}\delta_{\alpha\lambda}\delta_{js}\delta_{\beta\mu}+\delta_{i\alpha}\delta_{r\lambda}\delta_{j\beta}\delta_{s\mu}+\delta_{i\lambda}\delta_{r\alpha}\delta_{j\mu}\delta_{s\beta}\right]. \label{eq:O2av}
 \end{eqnarray}
 
 This immediately gives
 \begin{eqnarray}
 \overline{||\bm{y}(t)||^2}&=&\frac{1}{N}\tr A, \\
  \overline{||\bm{y}(t)||^4}&=&\frac{1}{N(N+2)}\left(2\tr A^2+(\tr A)^2\right),
 \end{eqnarray}
 which leads to the formula resembling the standard expression for variance, but rescaled by an additional factor
 \begin{equation}
 \mbox{var}(||\bm{y}(t)||^2) = \frac{2}{N+2}\left[\frac{1}{N}\tr A^2-\left(\frac{1}{N}\tr A\right)^2\right]. \label{eq:Variance}
 \end{equation}

\subsection{Randomness in the coupling matrix}
In general, functions of matrices and their transposes are difficult to handle, because they mix eigenvalues and eigenvectors in a non-uniform way. For example, the density of eigenvalues of $e^{Jt}$ is obtained by pushforward of the eigenvalue density of $J$, but these two matrices share the same eigenvectors. To deal with general functions of matrices, we resort to the matrix version of the Cauchy integration theorem, which allows us to reduce the problem to taking expectations of resolvents and their products for the price of contour integration. Let $f$ be an analytic function, then
\begin{equation}
f(X) = \frac{1}{2\pi i} \oint_{\gamma} \frac{f(z)dz}{z-X},
\end{equation}
where the contour $\gamma$ encircles counterclockwise the region occupied by all eigenvalues. The large $N$ limit of a random matrix is a certain noncommutative operator~\cite{MingoSpeicher}, but this representation is naturally generalized to the Dunford-Taylor integral~\cite{DunfordTaylor1,DunfordTaylor2}.

This allows us to write the following representation of the average squared norm
\begin{equation}
S(t) = \frac{e^{-2\mu t}}{(2\pi i)^2} \oint_{\gamma} e^{t(z_1+w_1)} W_1(z_1,w_1) dz_1 dw_1,
\end{equation}
where we denoted the correlation function
\begin{equation}
W_1(z_1,w_1) = \<\frac{1}{N}\tr \frac{1}{z_1-X}\frac{1}{w_1-X^{T}} \> = \< \frac{1}{N}\tr R_{X}(z_1)R_{X^T}(w_1)\>.
\end{equation}
To keep our notation concise, we denote the resolvent as $R_X(z) = (z-X)^{-1}$.
Analogously, for the variance we use the representation
\begin{equation}
\Sigma(t) = \frac{2}{N+2} \frac{e^{-4\mu t}}{(2\pi i )^4} \oint_{\gamma} e^{t(z_1+w_1+z_2+w_2)} \left[W_2(z_1,w_1,z_2,w_2)-W_1(z_1,w_1)W_1(z_2,w_2) \right] dz_1 dw_1 dz_2 dw_2,
\end{equation}
with $W_2(z_1,w_1,z_2,w_2) = \<\frac{1}{N}\tr R_X(z_1) R_{X^{T}}(w_1)R_X(z_2) R_{X^{T}}(w_2)\>$. %This formalism with the function $W_1$ has been successfully applied to dynamical systems~\cite{ChalkerMehlig, MartiBrunelOstojic, Erdos2019, NeriVivo}.

In the Rajan-Abbott model the coupling matrix $X$ is not fully random. It is decomposed into  random and deterministric parts $X=Y+M$. The deterministic part can be written as $M=\bm{um}^T$, with $\bm{m}^T=(m_1,m_2,\ldots,m_N)$ and $\bm{u}=\frac{1}{\sqrt{N}}(1,1,\ldots,1)^T$, thus it is a rank one update to the random part. We therefore use the Shermann-Morrison formula to write
\begin{equation}
R_{Y+M}(z)=(z-Y)^{-1}+\frac{(z-Y)^{-1}M(z-Y)^{-1}}{1-\bm{m}^T(z-Y)^{-1}\bm{u}}.
\end{equation}
We observe that the E/I balance condition implies $\bb{u}^T\bb{m}=0$ and $Y\bb{u}=0$. As a consequence also $YM=0$, thus  the resolvents of $Y+M$ and of $Y$ are related via
\begin{equation}
R_{Y+M}(z)=\left(1+\frac{M}{z}\right)R_{Y}(z).
\end{equation}
This in turn allows us to write correlation functions in terms of resolvents of $Y$:
\begin{eqnarray}
W^{(1)}_{Y+M}&=&\<\frac{1}{N}\tr\left[\left(1+\frac{M^T}{w_1}\right)\left(1+\frac{M}{z_1}\right)R_Y(z_1)R_{Y^T}(w_1)\right]\>, \label{eq:WYM1} \\
W^{(2)}_{Y+M}&=&\<\frac{1}{N}\tr \left[\left(1+\frac{M^T}{w_2}\right)\left(1+\frac{M}{z_1}\right)R_Y(z_1)R_{Y^T}(w_1) \left(1+\frac{M^T}{w_1}\right)\left(1+\frac{M}{z_2}\right) R_Y(z_2)R_{Y^T}(w_2) \right] \>. \label{eq:WYM2}
\end{eqnarray}

 The balance condition is imposed by subtracting the same number from each element in a row so that the sum within a row is zero. This is a small perturbation of a Ginibre matrix, which puts only $N$ constraints on $N^2$ random variables, so it is negligible in the large $N$ limit. We therefore assume that $Y$ has the same properties as Ginibre. In particular we assume that the probability density function is invariant under multiplication of an orthogonal matrix, $P(Y) = P(OYO^T)$. Then, the expectation over $O$ does not affect the expectation over $Y$, which allows us to write an alternative expression for the correlation functions, directly relating them to the ones for the Ginibre matrix
 \begin{eqnarray}
W^{(1)}_{Y+M}(z_1,w_1) & = & 
W_{Y}^{(1)}(z_1,w_1)\left(1+\frac{F}{z_1 w_1}\right), 
\\
 W^{(2)}_{Y+M}(z_1,w_1,z_2,w_2) & = & 
\frac{N}{(N-1)(N+2)}\left[ W^{(2)}_Y(z_1,w_1,z_2,w_2)\left((N+1)ab-\frac{c+d}{N}\right)+  \right. \nonumber 
\\ 
&&
 W^{(2)}_Y(z_1,w_1,w_2,z_2)\left(d\frac{N+1}{N}-ab-\frac{c}{N}\right)  +  \nonumber
 \\
&& \left. W^{(1)}_Y(z_1,w_1)W^{(1)}_Y(z_2,w_2) \left((N+1)c-Nab-d\right)
\right],
\label{eq:W2RA}
\end{eqnarray}

with 
\begin{eqnarray*}
a&=&1+\frac{F}{z_1w_2},\qquad \qquad  b=1+\frac{F}{z_2w_1}, \\
c&=&1+F\left(\frac{1}{z_1w_1}+\frac{1}{z_1w_2}+\frac{1}{z_2w_1}+\frac{1}{z_2w_2}\right)+\frac{NF^2}{z_1z_2w_1w_2}, \\
d&=&1+F\left(\frac{1}{z_1z_2}+\frac{1}{w_1w_2}+\frac{1}{z_2w_1}+\frac{1}{z_1w_2}\right)+\frac{NF^2}{z_1z_2w_1w_2}, 
\end{eqnarray*}
and $F=\frac{1}{N}\sum_{j=1}^{N}m_j^2$. As a consistency check of \eqref{eq:W2RA} one can verify that it reduces to $W^{(2)}_{Y}(z_1,w_1,z_2,w_2)$ for $F=0$.
 Since we already work with the large $N$ approximation, we keep only dominant terms in~\eqref{eq:W2RA}, which yield
 \begin{equation}
W^{(2)}_{Y+M}\sim \frac{NF^2}{z_1z_2w_1w_2}W^{(1)}_Y(z_1,w_1)W^{(1)}_Y(z_2,w_2). \label{eq:W2Fact}
\end{equation}

Finally, the correlation functions for the Ginibre read~\cite{EynardKristjansen,NowakProbing}
\begin{eqnarray}
W_{Y}^{(1)}(z_1,w_1) &=& \frac{1}{z_1 w_1-1}, \\
W_{Y}^{(2)}(z_1,w_1,z_2,w_2) &=& \frac{z_1z_2w_1w_2}{(z_1w_1-1)(z_1w_2-1)(z_2w_1-1)(z_2w_2-1)}.
\end{eqnarray}

\subsection{Evaluation of integrals} 
 To calculate the mean squared norm, we write
 \begin{equation}
 S(t) = \frac{e^{-2\mu t}}{(2\pi i)^2} \oint_{\gamma} \frac{e^{t(z_1+w_1)}}{z_1w_1-1}\left( 1+ \frac{F}{z_1w_1}\right) dz_1 dw_1.
 \end{equation}
 The integral over $w_1$ can be performed by the residue theorem. There are two residua at $w_1=0$ and $w_1=\frac{1}{z_1}$, yielding
 \begin{equation}
 S(t) = \frac{e^{-2\mu t}}{2\pi i}\oint_{\gamma} \frac{e^{tz_1}}{z_1}\left( (1+F)e^{\frac{t}{z_1}} - F\right)dz_1.
 \end{equation}
Using the representation of the modified Bessel function of the first kind 
\begin{equation}
I_{\nu}(t) = \frac{1}{2\pi i}\oint e^{\frac{t}{2}(z+1/z)}\frac{dz}{z^{\nu+1}}, \label{eq:BesselRepr}
\end{equation}
we arrive at
\begin{equation}
S(t) = e^{-2\mu t}\left((1+F)I_0(2t) - F\right).
\end{equation}
According to~\eqref{eq:W2Fact}  the contribution to the variance factorizes in the large $N$, thus we use the same technique to evaluate the integrals. Note also that $W_{Y+M}^{(2)} \sim N$, which cancels the $1/N$ factor in \eqref{eq:Variance}. Evaluating the integrals, we obtain
\begin{equation}
\Sigma(t) = 2F^2 e^{-4\mu t} (I_0(2t)-1)^2.
\end{equation}

Calculations of the variance for the pure Ginibre matrix are more involved, as integrals do not separate. We start with
\begin{equation}
\Sigma_{Gin}(t) = \frac{2}{N} \frac{e^{-4\mu t}}{(2\pi i)^4} \oint_{\gamma} e^{t(z_1+z_2+w_1+w_2)}\left(W^{(2)}_Y(z_1,w_1,z_2,w_2)-W_Y^{(1)}(z_1,w_1)W_Y^{(1)}(z_2,w_2)\right)  dz_1 dz_2 dw_1 dw_2 
\end{equation}
Integrations over $w_1$ and $w_2$ can be performed by residues, leading to
\begin{equation}
\Sigma_{Gin}(t)=\frac{2e^{-4\mu t}}{N(2\pi i)^2} \oint_{\gamma} dz_1 dz_2 \frac{e^{t(z_1+z_2)}}{z_1z_2} \frac{\left(e^{t/z_1}-e^{t/z_2}\right)\left(z_2^2 e^{t/z_1}-z_1^2 e^{t/z_2}\right)}{(z_1-z_2)^2}. \label{eq:VarGintegral}
\end{equation}
Note that the integrand is not singular at $z_2=z_1$, thus without loss of generality we assume that $|z_1|>|z_2|$. This allows us to expand
\begin{equation}
\frac{1}{(z_1-z_2)^2}=\frac{1}{z_1^2}\sum_{k=0}^{\infty}(k+1)\left(\frac{z_2}{z_1}\right)^k
\end{equation}
and rewrite \eqref{eq:VarGintegral} in the following form:
\begin{equation}
\Sigma_{Gin}(t)=\frac{2}{N}e^{-4\mu t}\sum_{k=0}^{\infty}(k+1)\left(A_k-B_k-C_k+D_k\right),
\end{equation}
with
\begin{eqnarray}
A_k&=&\frac{1}{(2\pi i)^2}\oint_{\gamma} dz_1 dz_2 e^{t(z_1+2/z_1)}e^{tz_2} \frac{z_2^{k+1}}{z_1^{k+3}}=0, \\
B_k&=&\frac{1}{(2\pi i)^2}\oint_{\gamma} dz_1 dz_2 e^{t(z_1+1/z_1)}e^{t(z_2+1/z_2)} \frac{z_2^{k+1}}{z_1^{k+3}}=I_{k+2}^2(2t), \\
C_k&=&\frac{1}{(2\pi i)^2}\oint_{\gamma} dz_1 dz_2 e^{t\left(z_1+\frac{1}{z_1}\right)}e^{t\left(z_2+\frac{1}{z_2}\right)}\frac{z_2^{k-1}}{z_1^{k+1}}=I_k^2(2t), \\
D_k&=&\frac{1}{(2\pi i)^2}\oint_{\gamma} dz_1 dz_2 e^{tz_1}e^{t\left(z_2+\frac{2}{z_2}\right)}\frac{z_2^{k-1}}{z_1^{k+1}}=\frac{t^k}{k!}2^{k/2}I_k(2\sqrt{2}t),
\end{eqnarray}
where we have used the Bessel function representation~\eqref{eq:BesselRepr}. Note that 
\begin{equation}
\sum_{k=0}^{\infty}(k+1)(B_k+C_k)=I_0^2(2t)+\sum_{k=0}^{\infty}2kI_k^2(2t)=I_0^2(2t)+2t I_1(2t)I_0(2t),
\end{equation}
where the last sum can be found in~\cite{Prudnikov}. To evaluate the sum of $D_k$, we use the series expansion of the Bessel function
\begin{equation}
\sum_{k=0}^{\infty}(k+1)D_k=\sum_{k,l=0}^{\infty}(k+1)\frac{(2t^2)^{k+l}}{k!l!(k+l)!}=I_0(4t)+tI_1(4t).
\end{equation}
Putting these results together, we obtain
\begin{equation}
\Sigma_{Gin}(t)=\frac{2}{N}e^{-4\mu t}\left[I_0(4t)+tI_1(4t)-2tI_1(2t)I_0(2t)-I_0^2(2t)\right].
\end{equation}

\section{Acknowledgments}
This work was supported by the Etiuda scholarship UMO2018/28/T/ST1/00470 from National Science Center and  TEAMNET POIR.04.04.00-
00-14DE/18-00 grant of the Foundation for Polish Science.
The author would like to thank Maciej A. Nowak, Jacek Grela, Izaak Neri, Pierpaolo Vivo and Dominik Suszalski for fruitful discussions.

\bibliography{refs}

\end{document}